\documentclass[aps,preprint]{revtex4}

\usepackage{graphicx}

\begin{document}

\date{\today}

\title{Resonances in  open quantum systems} 

\author{
Hichem Eleuch$^{1}$\footnote{email: hichemeleuch@tamu.edu} and 
Ingrid Rotter$^{2}$\footnote{email: rotter@pks.mpg.de}}

\address{
$^1$ Institute for Quantum Science and Engineering,
Texas A$\&$M University, College Station, Texas 77843, USA}
\address{
$^2$ Max Planck Institute for the Physics of Complex Systems,
D-01187 Dresden, Germany  }

\vspace*{1.5cm}

\begin{abstract}

The Hamilton operator of an open quantum system is non-Hermitian.
Its eigenvalues are, generally, 
complex and provide not only the energies but also the lifetimes 
of the states of the system. 
The states may couple via the common environment of scattering 
wavefunctions into which the system is embedded. This causes an  
{\it external mixing} (EM) of the states. Mathematically, 
EM is related to the existence of singular (the so-called exceptional) 
points (EPs).
The eigenfunctions of a non-Hermitian operator are biorthogonal,
in contrast to the orthogonal eigenfunctions of a Hermitian operator. 
A quantitative measure for the ratio between biorthogonality and 
orthogonality is the phase rigidity of the wavefunctions. 
At and near an EP, the phase rigidity takes its minimum value.
The lifetimes of two nearby eigenstates of a quantum system 
bifurcate under the influence of an EP. At the parameter value of 
maximum width
bifurcation, the phase rigidity approaches the value one, meaning that
the two eigenfunctions become orthogonal. However, the eigenfunctions
are externally mixed at this parameter value.
The S-matrix and therewith
the cross section do contain, in the one-channel case, almost no
information on the EM  of the states.
The situation is completely different in the case with two (or more)
channels where the resonance structure is strongly influenced by 
the EM of the states and interesting features of non-Hermitian
quantum physics are revealed. We provide numerical 
results for two and three nearby eigenstates of a non-Hermitian 
Hamilton operator which are embedded in one common continuum
and are influenced by two adjoining EPs. The results are discussed. 
They are of interest for an experimental test of the non-Hermitian 
quantum physics as well as for applications.

\end{abstract}

\pacs{\bf }
\maketitle

\section{Introduction}
\label{intr}

In  experiments \cite{yacobi,schuster,heiblum} 
on Aharonov-Bohm rings containing a quantum dot in one arm, 
both the phase and magnitude of the transmission amplitude
$T=|T|~e^{i\beta}$ of the dot can be extracted. The obtained results 
did not fit into the general understanding of the transmission process. 
As a  function of the plunger gate voltage $V_g$, a series of well-separated 
transmission peaks of rather similar width and height has been observed
and, according to expectations, the transmission phases $\beta(V_g)$ 
increase continuously by $\pi$ across every resonance. 
In contrast to expectations, however, 
$\beta$ always jumps sharply downwards by $\pi$  in each valley
between any two successive peaks. These jumps, called
{\it phase lapses in multi-level  systems}, were observed 
in a large succession of valleys for every  many-electron dot studied.
They have been discussed in many theoretical papers, 
including in a Focus Issue in New J. Phys. 2007. 

In spite of much effort, the experimental
results could not be explained in Hermitian quantum physics. 
Using the non-Hermitian formalism of the quantum physics, 
is was however possible, in 2009, to explain    \cite{muro}
convincingly the experimentally observed phase lapses
(see also the discussion of this problem in
Sect. 4.3.2  of the recent review \cite{ropp}). 
This example shows the meaning which non-Hermitian quantum physics 
{\it can}  have for the description of a concrete physical system 
that is open, in contrast to a closed (or almost closed) system that 
is well described in the framework of Hermitian quantum physics.

Another example which shows the meaning of a  non-Hermitian Hamilton
operator for  a concrete quantum system, is  
the description of laser-induced continuum structures in atoms 
more than 15 years ago
\cite{marost12,marost3}. In these  papers the motion of the complex 
eigenvalues of the non-Hermitian Hamiltonian 
is traced as a function of the field strength for different field
frequencies and atomic parameters. Level repulsion in
the complex plane is shown to occur at a critical field intensity.
With further increasing intensity, the complex energies move 
differently. This effect is called {\it resonance trapping} according
to similar results obtained earlier in nuclear physics \cite{ro91}. 

Recently, non-Hermitian Hamilton operators are used for the description 
or prediction of different phenomena in quantum physics,
e.g. \cite{moiseyev}. We mention here only a few of many examples
\cite{atabek,gilary,jaouadi}.    
 
A non-Hermitian Hamilton operator describing an open quantum system, 
may play an important role also 
in explaining well-known puzzles of quantum physics.  
The natural environment of a  localized quantum mechanical system 
is the  extended continuum of scattering wavefunctions in which
the system is embedded. This environment can be 
changed by means of external forces, however it can never be deleted.
It exists at all times and is completely independent of any observer.
For this reason, radioactive dating can be used in geologic studies. 

According to this statement, the properties of an open quantum 
system can be described by means of
two projection operators each of which is related to
one of the two parts of the function space. The localized part of 
the quantum system is basic for spectroscopic studies. 
Mathematically, the localized part of the open quantum system
is a subsystem that is related to another subsystem. 
The Hamiltonian of the (localized) system is 
therefore non-Hermitian while the Hamiltonian of the total system
consisting of the two subsystems, is Hermitian \cite{comment3}.

In the standard Hermitian description of a localized quantum system, 
the system is considered to be closed; the Hamiltonian is Hermitian; 
the eigenstates are discrete. 
Their decay is described by tunneling of particles into
the function space of scattering states into which the system 
is embedded. The tunneling time can be calculated. It is small and
could be measured experimentally only recently \cite{tunnel}. The
experimental results have shown that the tunneling time is
extremely short what does not correspond to the expectations of 
standard Hermitian quantum physics. They agree, however, with the
conclusions obtained when the system is considered to be an open
system described by a non-Hermitian Hamilton operator. In this case, 
the eigenvalues ${\cal E}_i$ of the system are complex and the 
lifetime of the states is given by Im(${\cal E}_i$). There is no need 
to consider any tunneling time.

In a similar manner, the problem of the  Schr\"odinger cat
does not exist when the system is considered to be an open quantum 
system. The price to be paid for this 
is  to  describe the system by a non-Hermitian Hamilton operator 
and to solve the involved mathematical problems.

Usually, the calculations with a non-Hermitian Hamiltonian  
give results for observable values of the quantum system which differ 
only little from those obtained with a Hermitian Hamiltonian,
especially in relation to the uncertainties involved in the
comparison with experimental data. There are however exceptions from
this rule. These exceptions   
arise from the mathematical existence of singular points.
One example are the so-called {\it exceptional points} (EPs)  which
are known in mathematics for many years \cite{kato}.  
Consider a family of operators of the form 
\begin{eqnarray}
\label{kato1}
T(\kappa)=T(0)+\kappa T'
\end{eqnarray}
where  $\kappa$ is a  scalar parameter,
 $T(0)$  is an unperturbed operator and
 $\kappa T'$  is a perturbation.
Kato \cite{kato} has shown that the  
number of eigenvalues of $T(\kappa)$ is independent of $\kappa$, as
expected, however
with the exception of some special values of $\kappa$. The
corresponding points in the parameter space are called
EPs. Here, (at least) two eigenvalues coalesce. An example is
\begin{eqnarray}
\label{kato2}
T(\kappa) = \left(
\begin{array}{cc}
1 & \kappa \\
\kappa & -1 
\end{array} \right) 
 \end{eqnarray} 
in which the two  values
$\kappa = + i$ and $\kappa = - i$ result in the
same eigenvalue $0$. 

Now the following questions arise: 
What is the behavior of the eigenfunctions of the non-Hermitian 
Hamilton operator  under the influence of an EP?
Can EPs be observed directly in experimental results?
These questions are answered only partly in the literature up to now
although their
influence onto  the dynamics of open quantum systems is
well known (for references see the review \cite{top}). 

The meaning of EPs is studied in literature during last about twenty
years, in classical as well as in quantum physics. We will not discuss
here the problems of classical physics. Instead we refer to the
excellent book \cite{kirillov} and to the collection of articles 
on spectral analysis, stability and bifurcation in modern nonlinear 
physical systems \cite{kir_pel}. Related problems are studied also in 
molecular physics, see the recent paper \cite{past} where the
references to older papers can be found. 
In quantum physics the problems are studied, unfortunately, 
in a confusing and often contradictory manner. We will not 
enumerate here the references to the different papers. 
They will rather be cited in those sections of the present paper 
in which they can be discussed consistently \cite{comment4}. 

The non-Hermiticity of the Hamiltonian is introduced mostly by adding
a non-Hermitian part to the Hermitian Hamiltonian that is known to
describe the system  quite well, for references see the review \cite{top}. 
It appears therefore more or less as a perturbation that is able to
describe the changes of the system properties under special
conditions, i.e. under the influence of an EP. Although this
treatment gives mostly reliable results, the question remains open  
what are the properties of a Hamiltonian which is really
non-Hermitian, i.e. when the non-Hermiticity does {\it not} appear 
as some perturbation.   

The aim of the present paper is to find an answer to 
this question in a mathematically exact manner, however by 
keeping in mind that points in the continuum are of measure zero and
cannot be observed directly. It is important therefore to point to
observable signatures of the EPs \cite{comment2} occurring  in
physical values, by means of which their existence can be proven. This is,
e.g., avoided level crossing  and formation of different time scales 
in the two-level case. Similar signatures exist in the three and 
more level cases. Most interesting is the 
so-called {\it external mixing (EM)} of the
states via the common continuum into which the system is embedded.
By definition, an EM of the states can occur {\it only} when the
system is open.  It is therefore one of the characteristic values of 
the non-Hermitian physics of open quantum systems \cite{comment3}.  

Many years ago, EM has been shown to play an important role
in the open quantum mechanical nuclear system (continuum shell model
in contrast to the standard shell model)  \cite{ro91}. 
Today we know that it
characterizes the main features of the influence of EPs
on the dynamics of an open quantum system, see the recent review
\cite{ropp} on experimental and theoretical results.
Experimentally, an example of EM has been provided 
a few years ago in a mesoscopic system. It
has been shown in \cite{bird1}  that two distinct quantum
states  are coupled through a common continuum. 
In a further experiment, the authors were
able to show that EM  survives even under conditions of strongly
nonequilibrium transport in the system \cite{bird2}.

The present paper is organized as follows. First we consider the 
Hamiltonian which describes the properties of an open
quantum system. By definition, an
open quantum system is localized in space and embedded
in the continuum of scattering wavefunctions
due to which the states of the system become resonance states
and have, generally, a finite lifetime.
This Hamiltonian is non-Hermitian. 
In Sects. \ref{eigf2} and \ref{eigf3}, respectively,
we consider the eigenvalues and
eigenfunctions of a $2\times 2$  and of a $3\times 3$ 
non-Hermitian Hamilton operator. 
The eigenstates are coupled via one common 
continuum and show  the typical EM.
The eigenfunctions of a non-Hermitian Hamilton operator
are biorthogonal,  and  their phases are
not rigid in approaching an EP. 

We show in the next section \ref{nonl}  that the
EPs cause nonlinear effects in an open quantum system.
They can be traced best in the resonance structure of the
scattering cross section under the condition that 
it is influenced by two adjoining EPs.
The further results given in  Sect. \ref{smatr1c} show that,
in the one-channel case,  the resonance structure of the cross section 
is almost independent of  the EM. 
It is therefore impossible to reason, in this case, the
existence of EPs from a study of the resonance structure of the cross
section. The situation is another one in the
case with two (or more) channels as discussed in Sect. \ref{smatr2c}. 
Here,  EPs and EM cause interesting observable effects.      

In Sect. \ref{disc}, we summarize and discuss the results obtained 
in the present paper.  We conclude the paper in
the last section \ref{concl} with some  general remarks on EPs and,
above all, on the eigenfunctions of a non-Hermitian Hamilton
operator. By doing this, we hope to stimulate experimental studies in
order to prove, on the one
hand, the theoretical results, and to use, on the other hand, the 
rich possibilities they provide for applications.

\section{Eigenvalues and eigenfunctions of a $2\times 2$
non-Hermitian Hamiltonian}
\label{eigf2}

Let us consider the $2\times 2$ non-Hermitian matrix
\begin{eqnarray}
\label{ham2}
{\cal H}^{(2)} = 
\left( \begin{array}{cc}
\varepsilon_{1} \equiv e_1 + \frac{i}{2} \gamma_1  & ~~~~\omega   \\
\omega & ~~~~\varepsilon_{2} \equiv e_2 + \frac{i}{2} \gamma_2   \\
\end{array} \right) \; .
\end{eqnarray}
Here, $\varepsilon_i$ are the complex eigenvalues of the basic 
non-Hermitian operator \cite{comment1}.
The $\omega$ stand for the coupling matrix elements of the two
states via the common environment \cite{comment3}. Their mathematical
expression is derived  in Sect. 3 of \cite{top}. They
are complex where Re($\omega$) is the principal value integral 
and Im($\omega$) is the residuum \cite{top}.
The imaginary part is responsible for coherent processes 
occurring in the system, while the real part contains
decoherences. The non-Hermitian matrix 
\begin{eqnarray}
\label{ham0}
{\cal H}_0^{(2)} = 
\left( \begin{array}{cc}
\varepsilon_{1} \equiv e_1 + \frac{i}{2} \gamma_1  & 0   \\
0 & ~~~~\varepsilon_{2} \equiv e_2 + \frac{i}{2} \gamma_2   \\
\end{array} \right) 
\end{eqnarray}
describes the system without any
mixing of its states via the environment. In other words, $\omega =0$ 
corresponds to vanishing EM of the eigenstates.

In this paper, our main interest is in the effects caused by
$\omega$. Most visible are the 
changes in the widths of the states: the 
original widths $\gamma_i$ of the states 
turn into the widths $\Gamma_i$ of the eigenstates of ${\cal H}^{(2)}$
due to $\omega \ne 0$.

\subsection{Eigenvalues of ${\cal H}^{(2)}$ }
\label{eigv}

The eigenvalues ${\cal E}_i \equiv E_i + \frac{1}{2} \Gamma_i$ 
of ${\cal H}^{(2)}$ are, generally, complex:
\begin{eqnarray}
\label{eig1}
{\cal E}_{1,2} \equiv E_{1,2} + \frac{i}{2} \Gamma_{1,2} = 
\frac{\varepsilon_1 + \varepsilon_2}{2} \pm Z 
\end{eqnarray}
with
\begin{eqnarray}
\label{eig2}
Z \equiv \frac{1}{2} \sqrt{(\varepsilon_1 - \varepsilon_2)^2 + 4
  \omega^2}
\; .
\end{eqnarray}
Here,  $E_i$ is the energy and $\Gamma_i$  the width of the eigenstate $i$. 

The properties of the ${\cal E}_i$ trajectories as a function of a
certain parameter are well known. They contain
\begin{description}
\item[{\it ~Level repulsion:}]
two states repel each other in accordance with Re$(Z)$;
\item[{\it ~Width bifurcation:}]
the widths of two states bifurcate in accordance with Im$(Z)$;
\item[{\it ~Avoided level crossing:}]
two discrete (or narrow resonance) states avoid crossing
\cite{landau,zener}
because $ (\varepsilon_1 - \varepsilon_2)^2 + 4 \omega^2
>0$ and therefore always $Z\ne 0$;
\item[{\it ~Appearance of an EP:}]
two states cross when $Z=0$. 
\end{description}

Altogether, the crossing scenario that is caused by an EP
in non-Hermitian quantum physics,
with, generally, complex eigenvalues ${\cal E}_i \equiv
E_i + \frac{1}{2} \Gamma_i$ of the Hamiltonian,
needs to be considered in terms 
of a combined behavior of energy ($E_i$) and width 
($\Gamma_i$) trajectories of the two states $i=1,~2$.
A level repulsion will generally appear in the  
Re$({\cal E}_i) = E_i $ trajectories together with a free crossing 
of the  Im$({\cal E}_i) \propto \Gamma_i $
trajectories; while a bifurcation of the widths $\Gamma_i$ 
is accompanied, generally, by a free crossing of the energy
trajectories  $E_i $. The last case
is illustrated, some years ago, in Figs. 1 to 3 
of the paper \cite{mudiisro}.
Sometimes, the crossing phenomenon in non-Hermitian quantum physics 
is called  {\it avoided level crossing in the complex plane}, 
e.g. \cite{burosa1}.  In the present paper we use the 
term {\it avoided level crossing} or {\it level repulsion} according
to the standard definition for  
the  Re$({\cal E}_i) = E_i$ trajectories;
while the term {\it width bifurcation} is used for 
the corresponding phenomenon appearing in the Im$({\cal E}_i) \propto 
\Gamma_i$ trajectories. We underline once more: {\it both phenomena are 
combined} in non-Hermitian quantum physics. 

In \cite{pra93}, the case with equal widths $\gamma_1 = \gamma_2$ of 
the two states and with imaginary coupling $\omega = i 
\,\omega_0$ is solved analytically. As a result, two EPs appear, see 
Eqs. (14)-(16) and
Fig. 1.a-d in \cite{pra93}. Between the two EPs, the widths 
bifurcate up to a maximum value. In the present paper, we 
consider complex  $\omega_i$ where only one EP can be seen clearly. 
Nevertheless,  also the second EP has some influence onto the 
dynamical properties of the system (see the numerical
results given in the present paper).    

When $\omega = 0$, the energies $\varepsilon_i$ vary smoothly as a
function of any parameter. According to (\ref{eig1}) and
(\ref{eig2}), $Z= \pm \frac{1}{2}(\varepsilon_1 - \varepsilon_2) $ and 
${\cal E}_{1,2} \to \varepsilon_{1,2}$ in this case. This means that no EP 
can be related to the Hamiltonian   ${\cal H}_0^{(2)}$.

\subsection{Eigenfunctions  of ${\cal H}^{(2)}$   }
\label{eigfu}

The properties of the eigenfunctions $\Phi_i$ of a non-Hermitian
operator are less known. They are the following.

\begin{description}
\item[{\it ~~Biorthogonality:}]
The  eigenfunctions and eigenvalues of every Hamilton operator 
have to fulfill the two conditions 
\begin{eqnarray}
\label{eif1}
{\cal H} |\Phi_i\rangle =  {\cal E}_i|\Phi_i\rangle \hspace*{1cm}
\langle \Psi_i|{\cal H} = {\cal E}_i \langle  \Psi_i|\; .
\end{eqnarray}
A Hermitian operator has real eigenvalues such that  
$\langle \Psi_i| = \langle \Phi_i|$ in this case.\\
The eigenvalues of a  non-Hermitian operator
are generally complex such that the left and right eigenfunctions
differ from one another,
$\langle \Psi_i| \ne \langle \Phi_i|$.\\
This is valid also for the eigenvalues and eigenfunctions of
the two symmetric operators 
${\cal H}^{(2)}$ and  ${\cal H}_0^{(2)}$. In this case, the
relation between the left and right eigenfunctions 
is given by \cite{mudiisro,savin1,savin2}
\begin{eqnarray}
\label{eif2}
\langle \Psi_i| = \langle \Phi_i^*|\; .
\end{eqnarray}

\item[{\it ~~Normalization:}]
In the case of a Hermitian operator,
$\langle \Phi_i|\Phi_j\rangle$ is real and the eigenfunctions are usually
normalized to $\langle \Phi_i|\Phi_j\rangle =1$.\\
To smoothly describe the transition from a closed system 
with discrete states to a weakly open one with narrow resonance states
(described by ${\cal H}^{(2)}$), it is meaningful to use the normalization
\begin{eqnarray}
\label{eif3}
\langle \Phi_i^*|\Phi_j\rangle = \delta_{ij} 
\end{eqnarray}
for the eigenfunctions. The value 
$\langle \Phi_i^*|\Phi_j\rangle \equiv (\Phi_i|\Phi_j) $ is 
however complex such that
the phases of the two eigenfunctions $\Phi_{1,2}$ relative 
to one another cannot be rigid. They are rather parameter dependent
since $\langle \Phi_i^*|\Phi_j\rangle$ has to be real,
according to (\ref{eif3}), for every parameter value.

It follows from (\ref{eif3}),  that the values of the
standard expressions  are changed \cite{top},
\begin{eqnarray}
\label{eif4}
 \langle\Phi_i|\Phi_i\rangle  =  
{\rm Re}~(\langle\Phi_i|\Phi_i\rangle) ~; \quad
A_i \equiv \langle\Phi_i|\Phi_i\rangle \ge 1 
\end{eqnarray}
\vspace*{-1.2cm}
\begin{eqnarray}
\label{eif5}
\nonumber
\langle\Phi_i|\Phi_{j\ne i}\rangle  = 
i ~{\rm Im}~(\langle\Phi_i|\Phi_{j \ne i}\rangle) =
-\langle\Phi_{j \ne i}|\Phi_i\rangle 
\end{eqnarray}
\vspace*{-1.4cm}
\begin{eqnarray} 
\label{eif6}
 |B_i^j|  \equiv |\langle \Phi_i | \Phi_{j \ne i}| ~\ge ~0  
\end{eqnarray}

\item[{\it ~~Phase rigidity:}]

The phase rigidity is a quantitative measure for the biorthogonality 
of the eigenfunctions. It is defined by \cite{top}
\begin{eqnarray}
\label{eif7}
r_k ~\equiv ~\frac{\langle \Phi_k^* | \Phi_k \rangle}{\langle \Phi_k 
| \Phi_k \rangle} ~= ~A_k^{-1} 
\end{eqnarray}
by taking into account the normalization (\ref{eif3}).

In Hermitian systems, the eigenfunctions are  orthogonal and
$r_k=1$. 

In systems with well-separated resonance states, it follows
$r_k\approx 1$; however it is never $r_k= 1$ 
\cite{top,savin1,savin2}.   
Hermitian quantum physics is, in this case. a reasonable
approximation for the description of the open quantum system.

In approaching an exceptional point, it follows  $r_k \to  0$
\cite{top}.

The phase rigidity is experimentally studied on microwave billiards 
\cite{richter2}. The variation of $r_k $
in approaching the EP is found, indeed. 
The experimental result agrees with the relation (\ref{sec8}) 
discussed below \cite{comment4}.

Our calculations show an interesting unexpected property
for two nearby states with similar values of their widths
$\gamma_i$ \cite{epj1,pra93}: 
$r_k\approx 1$ at maximum width bifurcation.
These results will be discussed below in detail.
An analog result  is found
for two nearby states with  level repulsion 
which is caused by an EP \cite{pra93}.

\item[{\it ~~Mixing of the eigenfunctions via the environment (EM):}]

The Schr\"odinger equation for the basic wave functions 
$\Phi_i^0 $ with the Hamiltonian (\ref{ham0}) is
\begin{eqnarray}
\label{eif8}
({\cal H}_0^{(2)}  - \varepsilon_i) ~| \Phi_i^0 \rangle  =  0 
\end{eqnarray}
while the Schr\"odinger equation with the full Hamiltonian 
(\ref{ham2}) reads 
\begin{eqnarray}
\label{eif9}
({\cal H}^{(2)}  - {\cal E}_i) ~| \Phi_i \rangle  =  0 \; .
\end{eqnarray}
Eq. (\ref{eif9}) can be rewritten in a Schr\"odinger equation with
source term,  
\begin{eqnarray}
\label{eif11}
({\cal H}^{(2)}_0  - {\cal E}_i) ~| \Phi_i \rangle  = -
\left(
\begin{array}{cc}
0 & \omega \\
\omega & 0 
\end{array} \right) |\Phi_i \rangle \; . 
\end{eqnarray}
Now, we can use the standard representation of the $\Phi_i$  in the 
$\{ \Phi_n^0 \}$ 
\begin{equation}
\label{eif12}
\Phi_i=\sum \, b_{ij} ~\Phi_j^0 \; ;
 \quad \quad b_{ij} = \langle \Phi_j^{0 *} | \Phi_i\rangle  
\end{equation}
under the condition that the $b_{ij}$ 
are normalized by $\sum_j (b_{ij})^2 = 1$, i.e.
\begin{eqnarray}
\label{eif13}
\sum_j (b_{ij})^2
  = {\rm Re} [\sum_j(b_{ij} )^2]
 = \sum_j \{[{\rm Re} (b_{ij})]^2 - [{\rm Im} (b_{ij})]^2\}
 =1 \; .
\end{eqnarray}
We are interested in the probability of EM which is defined by
\begin{eqnarray}
\label{eif14}
\sum_j |b_{ij}|^2
 = \sum_j \{[{\rm Re} (b_{ij})]^2 + [{\rm Im} (b_{ij})]^2\} \; . 
\end{eqnarray}
From (\ref{eif13}) and  (\ref{eif14})  follows
\begin{eqnarray}
\label{eif15}
\sum_j |b_{ij}|^2 \ge 1 \; .
\end{eqnarray}
In the neighborhood of an EP, $\sum_j |b_{ij}|^2 \gg 1$; and 
$\sum_j |b_{ij}|^2 \to \infty $ 
in approaching an EP \cite{epj1}.\\
When the  maximum width bifurcation (or level repulsion) is
parametrically reached, the eigenfunctions $\Phi_i$ 
are almost orthogonal, however  EM contained in the wavefunctions  
of the eigenstates, is strong \cite{epj1}.\\

\item[{\it ~~Eigenfunctions of ${\cal H}^{(2)}$ at an EP:}] 
According to analytical and numerical results 
\cite{ro01,marost3,gurosa,berggren}, it is
\begin{eqnarray}
\label{sec8}
\Phi_1^{\rm cr} \to ~\pm ~i~\Phi_2^{\rm cr} \; ;
\quad \qquad \Phi_2^{\rm cr} \to
~\mp ~i~\Phi_1^{\rm cr}   
\end{eqnarray}  
where $\Phi_i^{\rm cr}$ are the eigenfunctions at an EP.
The EP is however a point in the continuum of scattering wavefunctions
and is therefore of measure zero. 
Hints to the existence of an  EP can be found in observable
values. These are, above all,  avoided level crossing and width
bifurcation which both are caused by an EP.

We mention here that the relations (\ref{sec8}) are in agreement with
experimental results obtained on microwave billiards
\cite{richter1}. 
These nice results \cite{comment4} are confirmed 
independently from one another
by different authors, e.g. \cite{ro01,marost3,gurosa,berggren}.

\end{description}

The eigenfunction $\Phi_i$ of the non-Hermitian Hamilton operator 
$\cal H$ is the main part of the wavefunction of the resonance state
$i$ inside the localized part of the system. The wavefunction of the
resonance state including its tail is given in Eq. (42) in \cite{top}.

\subsection{Numerical results}
\label{num2}

\begin{figure}[ht]
\begin{center}
\includegraphics[width=10.cm,height=14cm]{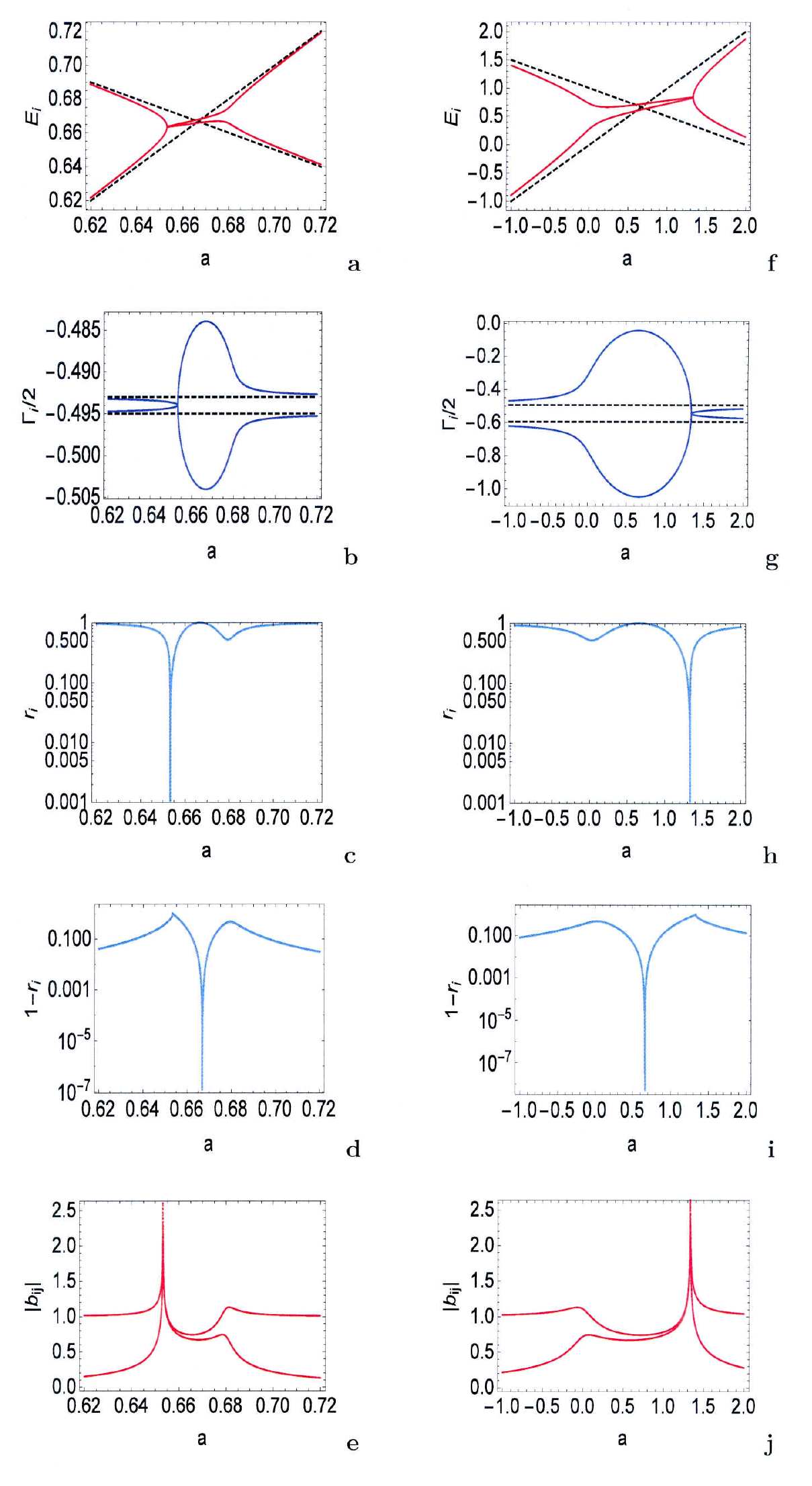} 
\vspace*{-.7cm} 
\end{center}
\caption{
\footnotesize{
Eigenvalues ${\cal E}_i \equiv E_i +\frac{1}{2}\Gamma_i$; phase
rigidity $r_i$ and $1-r_i$;  mixing $|b_{ij}|$ of the
eigenfunctions $\Phi_i$ of the Hamiltonian ${\cal H}^{(2)}$ as a
function of $a$. The value $\omega$ is independent of $a$: $\omega =
0.01(i+\frac{1}{10})$ (left) and $\omega = 0.5 (i+\frac{1}{10})$
(right). 
Parameters: 
$e_1=1-a/2; ~~e_2=a;
 ~~\gamma_1/2=-0.495; ~~\gamma_2/2=-0.493$ (left) and 
$e_1=1-a/2; ~~e_2=a; ~~\gamma_1/2=-0.495; ~~\gamma_2/2=-0.595$
(right). The dotted lines in (a,b,f,g)
show $e_i$ and $\gamma_i/2$, respectively.
}
}
\label{fig1}
\end{figure}

We refer to the analytical results obtained and discussed in \cite{pra93}
for the eigenvalues and eigenfunctions of $N=2$ states. In the 
analytical studies  $\omega$ is assumed to be either real or imaginary 
what is, of course, 
seldom realized in realistic systems. Nevertheless, the results of 
these studies provide some 
insight into the basic features of the eigenvalues and eigenfunctions 
of a non-Hermitian operator, above all near to an EP.

In Fig. \ref{fig1}, we show numerical results obtained for systems
under more realistic conditions in which $\omega$ is complex.
The energies $e_i$ are parameter dependent while the $\gamma_i$
as well as the $\omega$ are parameter independent. The difference 
between the widths $\gamma_i$ of the two states as well as 
$\omega$ are chosen in such a manner that an EP occurs.
In both cases, the phase rigidity $r_i$ approaches zero at the EP; 
and is near to one at maximum width bifurcation. The mixing 
$|b_{ij}|$ of the two eigenfunctions increases limitless in 
approaching the EP; and is finite, 
$|b_{ij}| < 1$, in the parameter region of the maximum width bifurcation. 
The hint to a second EP can be seen in the eigenvalues as well as 
in the eigenfunctions at, respectively, $a=0.68$ in Fig. \ref{fig1} left 
and  $a=0.0$ in Fig. \ref{fig1} right.   

Similar results are obtained when the energies $e_i$
(as well as the $\omega$) are chosen to be  
parameter independent while the $\gamma_i$ are parameter dependent,
see e.g.  \cite{pra93}. 

In our calculations, we choose the coupling strength $\omega$ between
system and environment parameter independent in order to exclude 
formally its influence onto the dynamics of the open quantum system. 
This allows us to fix the role of nonlinear processes.

To summarize the results of Fig. \ref{fig1}, we state the following. 

\begin{description}

\item[~~~~{\it Phase rigidity and mixing of the eigenfunctions 
in approaching an EP}] 
\begin{eqnarray}
\label{sec9}
r_i \to 0 
~~;~~~ |b_{ij}| \to \infty \; ,
\end{eqnarray}
in general
\begin{eqnarray}
\label{sec11}
1 > r_i \ge 0 ~~;~~~ |b_{ij}| > 1 \; .
\end{eqnarray}

\item[~~~~{\it Phase rigidity and mixing of the wavefunctions between two EPs
}]
\begin{eqnarray}
\label{sec12}
r_i \to 1 ~~;~~~ |b_{ij}| < 1
\end{eqnarray} 
in approaching maximum width bifurcation. 
In the analytically solvable case with imaginary $\omega$ it is \cite{epj1}
\begin{eqnarray}
\label{sec13}
|b_{ij}| \approx 0.7
\end{eqnarray}
meaning that the eigenfunctions 
$\Phi_i$ are almost orthogonal 
and strongly mixed in the set of 
basic wavefunctions \{$\Phi_k^{0}$\}
in approaching maximum width bifurcation.
\end{description}
 
Similar results are obtained when Re$(\omega) \gg {\rm Im}(\omega)$. 
The difference  to the results shown in Fig. \ref{fig1} is that now 
level repulsion is the main effect caused by the EP \cite{epj1}.

We remark here that the 
evolution from $r_k=0$ at the EP to $r_k\approx 1$ at the maximum 
width bifurcation
is driven exclusively by the
 nonlinear source term of the Schr\"odinger equation (see
 Sect. \ref{nonl}) 
since $\omega = const $  in our calculations.

When $\omega = 0$, it is  ${\cal E}_i = \varepsilon_i$. In this case, 
there are no EPs as mentioned in Sect. \ref{eigv}.

\section{Eigenvalues and eigenfunctions of a $3\times 3$
non-Hermitian Hamiltonian}
\label{eigf3}

Let us consider the Hamiltonian
\begin{eqnarray}
\label{nxn1}
{\cal H}^{(N)} = 
\left( \begin{array}{cccc}
\varepsilon_{1}  & \omega_{12} & \ldots &\omega_{1N}   \\
\omega_{21} & \varepsilon_{2}  &  \ldots & \omega_{2N}\\
\vdots     & \vdots &             \ddots&   \vdots \\
\omega_{N1} & \omega_{N2}       &    \ldots   &  \varepsilon_{N} \\
\end{array} \right)
\end{eqnarray}
where
$\varepsilon_{i} \equiv e_i + i/2~\gamma_i$ are the
energies and widths of the $N$ states;
$\omega_{i\,k\ne i}$
are the complex coupling matrix elements of the states $i$ and $k$ via 
the common environment; and the $\omega_{i\,k=i}$ denote   
the  selfenergy of the state $i$ which is  
mostly assumed to be included in the $\varepsilon_{i}$ in our 
calculations.

The values $\omega_{ik}$ for different $i$ and $k$ differ usually 
from one another. It is however a 
well-known fact from numerical calculations \cite{top}, that a resonance 
state becomes trapped by another 
nearby state  when its width is somewhat smaller than that of the 
nearby state. Finally the widths of most 
relatively short-lived states of the system are similar to one
another. These states determine the evolution of the system.

\subsection{Numerical results for $N=3$}
\label{num3}

\begin{figure}[ht]
\begin{center}
\includegraphics[width=10.cm,height=14.cm]{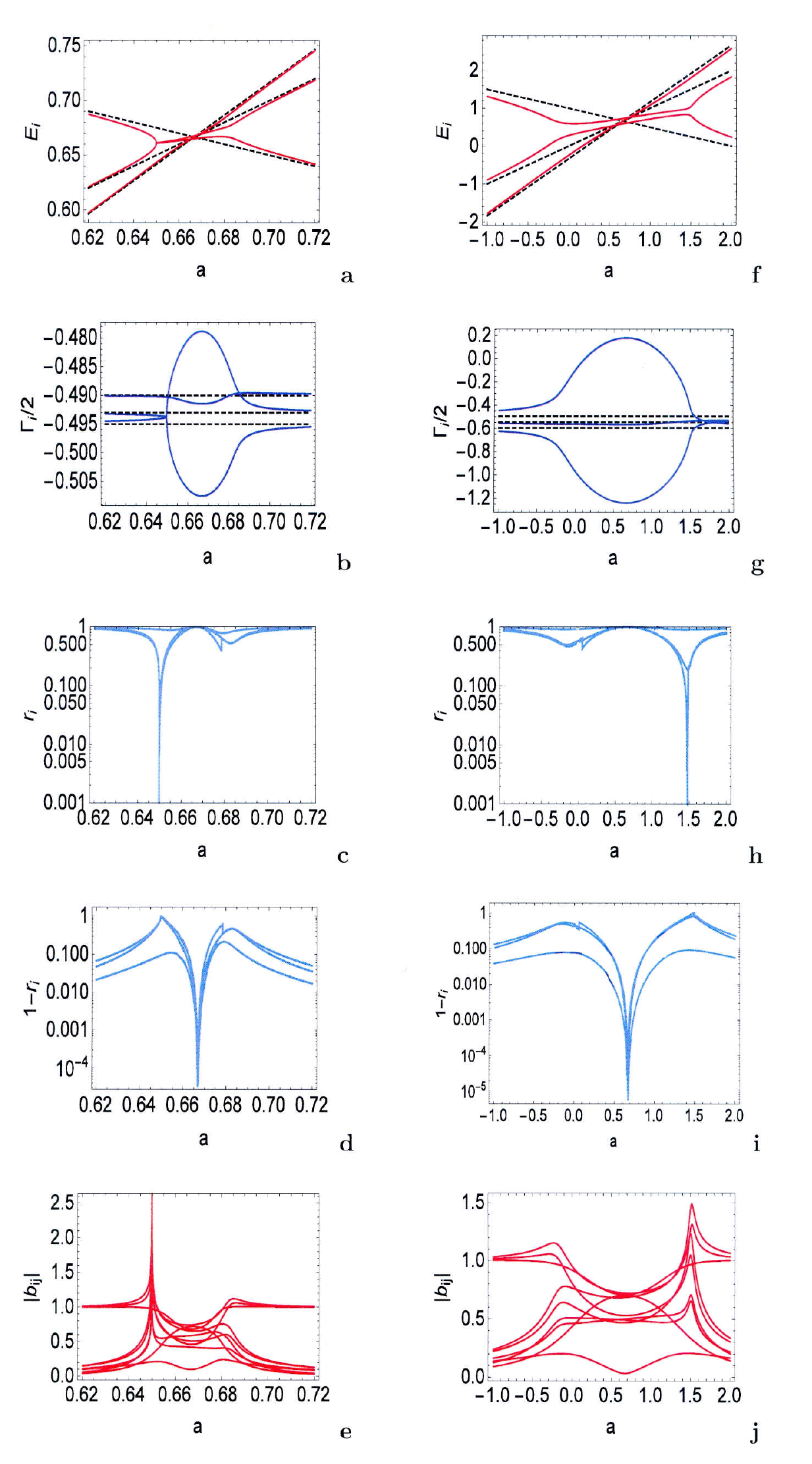} 
\vspace*{-.7cm}
\end{center}
\caption{
\footnotesize{
Eigenvalues ${\cal E}_i \equiv E_i +\frac{1}{2}\Gamma_i$; phase
rigidity $r_i$ and $1-r_i$;  mixing $|b_{ij}|$ of the
eigenfunctions $\Phi_i$ of the Hamiltonian ${\cal H}^{(3)}$ as a
function of $a$. The value $\omega$ is independent of $a$: $\omega =
0.01(i+\frac{1}{10})$ (left) and $\omega = 0.5 (i+\frac{1}{10})$
(right). Parameters: 
$e_1=1-a/2; ~~e_2=a; ~~e_3=-1/3+1.5~a;
 ~~\gamma_1/2=-0.495; ~~\gamma_2/2=-0.493;
~~\gamma_3/2=-0.49$ (left) and 
$e_1=1-a/2; ~~e_2=a; ~~e_3=-1/3+1.5~a;
 ~~\gamma_1/2=-0.495; ~~\gamma_2/2=-0.595
~~\gamma_3/2=-0.545$ (right). The dotted lines in (a,b,f,g)
show $e_i$ and $\gamma_i/2$, respectively.
}
}
\label{fig2}
\end{figure}

In Fig. \ref{fig2} we show the numerical results
obtained for the eigenvalues and eigenfunctions of $N=3$ states by using 
parameters similar to those for $N=2$ states in Fig. \ref{fig1}. 
The $\omega_{ij} \equiv \omega$ are
chosen to be equal for the different $i$ and $j$.
Above all, they are parameter independent, 
similar as the corresponding $\omega$ in Fig. \ref{fig1}. 
 
The comparison of Figs. \ref{fig1} and \ref{fig2}   shows that the 
main features of the eigenvalues and eigenfunctions are the same 
for $N=2$ and $N=3$.
The eigenvalues repel each other in energy and their widths 
bifurcate under the influence of an EP; the phase 
rigidities approach zero and the mixing of the wavefunctions 
increases limitless at and near to an 
EP; the phase rigidities approach the value one and the corresponding 
almost orthogonal wavefunctions are mixed
when the width bifurcation is maximum.  These effects are enhanced 
when $N=3$ as compared to those occurring when $N=2$.

As in the two-level case, similar results are obtained 
when Re$(\omega) \gg {\rm Im}(\omega)$. In this case, level repulsion 
is the main effect caused by the EP \cite{epj2}.

\subsection{Third-order exceptional points}
\label{third}

Hints to third-order EPs (at which three eigenvalues coalesce at one
parameter value) cannot be found in Fig. \ref{fig2}. The
reason is that every EP is a point in 
the continuum (with measure zero) which can be identified only by its 
influence onto observable values in a finite parameter range around it.   
Furthermore, a third-order EP occurring in the system without any EM
of its states via the environment, is shielded due to EM in a
realistic system. It can therefore not be observed in an open quantum
system.

According to the numerical results shown in Fig. \ref{fig2}, 
we see several second-order EPs in a critical 
parameter region around the value at which the 
conditions for a third-order EP are mathematically fulfilled. 
The observable effect caused by a third-order EP,
is  some clustering 
of second-order EPs which occurs in a finite 
parameter range around the value at which the third-order EP is 
mathematically expected. This fact is discussed in detail in \cite{pra93}.

These results show the differences between a formal-mathematical
result and
effects that can really be observed in a physical system. The point 
is that two states that cross at an EP
lose (due to the EM of the states) 
their individual character
in a finite parameter range around the EP; and  
the areas of influence of various
second-order EPs overlap. In this manner, they 
amplify, collectively, their impact onto physical values with the 
result that, e.g., a third-order EP
is shielded in a physical system.

\section{Schr\"odinger equation with nonlinear source term} 
\label{nonl}

The Schr\"odinger equation (\ref{eif9}) 
can be rewritten in the Schr\"odinger equation (\ref{eif11})
with source term. In this equation, 
the coupling $\omega$ of the  states  $i$ and ${j\ne i}$ 
via the common environment of scattering wavefunctions (EM)
is contained in the source term.

The source term is nonlinear \cite{top}
\begin{eqnarray}
\label{16}
({\cal H}_0^{(2)}  - {\cal E}_i) ~| \Phi_i \rangle  = 
\sum_{k=1,2} \langle
\Phi_k|W|\Phi_i\rangle \sum_{m=1,2} \langle \Phi_k |\Phi_m\rangle 
|\Phi_m\rangle 
\end{eqnarray}
since $\langle \Phi_k |\Phi_m\rangle \ne 1$ for $k= m$ and 
$\langle \Phi_k |\Phi_m\rangle \ne 0$ for $k\ne m$, see 
Eqs. (\ref{eif4}) and (\ref{eif6}).
In (\ref{16}) the definition $W \equiv 
-\left(
\begin{array}{cc}
0 & \omega \\
\omega & 0 
\end{array} \right)$
is used for convenience.

The most important part of the nonlinear contributions is contained in 
\begin{eqnarray}
\label{17}
({\cal H}_0^{(2)}  - {\cal E}_n) ~| \Phi_n \rangle =
\langle \Phi_n|W|\Phi_n\rangle ~|\Phi_n|^2 ~|\Phi_n\rangle \; . 
\end{eqnarray}
Far from an EP, the source term is (almost) linear since
$\langle \Phi_k|\Phi_{k }\rangle \to 1$ and
$\langle \Phi_k|\Phi_{l\ne k }\rangle = - 
\langle \Phi_{l \ne k  }|\Phi_{k}\rangle \to 0$. 
Near to an EP however, the source term is  nonlinear since
$\langle \Phi_k|\Phi_{k }\rangle \ne 1$ and
$\langle \Phi_k|\Phi_{l\ne k }\rangle = - 
\langle \Phi_{l \ne k  }|\Phi_{k}\rangle \ne 0$.

Due to the EM involved in the source term, the eigenfunctions $\Phi_i$ 
and  eigenvalues ${\cal E}_i$  
of ${\cal H}^{(2)}$ contain global features.
The environment of an open quantum system is
the continuum of scattering wavefunctions which
has an infinite number of degrees of freedom. It may cause therefore, 
among others, a dynamical phase transition \cite{top,epj2}. 
The transition is non-adiabatic \cite{top,ropp,epj2}.

In order to illustrate the nonlinear effects involved in the source 
term of the Schr\"odinger equation (\ref{16}) let us consider, as an 
example, the  resonance part of the $S$ matrix
from which the resonance structure 
of the cross section can be calculated, 
\begin{eqnarray}
\label{cro}
\sigma (E) \propto |1-S(E)|^2 \; .
\end{eqnarray}
A unitary representation of the resonance part of the 
$S$ matrix in the case of two resonance states coupled to a 
common continuum of scattering wavefunctions reads \cite{ro03} 
\begin{eqnarray}
\label{sm4}
S = \frac{(E-E_1-\frac{i}{2}\Gamma_1)~(E-E_2-\frac{i}{2}\Gamma_2)}{(E-E_1+
\frac{i}{2}\Gamma_1)~(E-E_2+\frac{i}{2}\Gamma_2)}\; .
\end{eqnarray}
Here, the influence of the EPs onto the cross section is contained 
in the eigenvalues 
${\cal{E}}_i = E_i + i/2~\Gamma_i$. The expression (\ref{sm4})
allows us therefore to receive  reliable results  also when the phase 
rigidity is reduced, $r_k < 1$.

The expression (\ref{sm4}) can be used in order to derive analytically 
an expression for the
resonance structure of the S-matrix at an EP
\cite{ro03},  
\begin{eqnarray}
\label{sm5}
S = 1-2i\frac{\Gamma_d}{E-E_d+\frac{i}{2}\Gamma_d}-
\frac{\Gamma_d^2}{(E-E_d+\frac{i}{2}\Gamma_d)^2} 
\end{eqnarray}
where $E_1=E_2\equiv E_d$ and
$\Gamma_1=\Gamma_2\equiv \Gamma_d$. As a result of interferences,  
this expression consists of three terms, one of which is   
explicitly nonlinear. The resonance structure (\ref{sm5}) shows 
two bumps approximately at the energies $\varepsilon_i$ of the 
two resonance states, and an interference minimum between them.
This structure resembles that of two more or less isolated 
resonances the energies of which are  $\varepsilon_1$ and $\varepsilon_2$.  

Many years ago, the resonance structure of the cross section with 
two resonance states is calculated  as a function of the coupling
strength 
between system and environment \cite{mudiisro}. 
These calculations are performed by using the standard expression 
for the S-matrix with the energies $\varepsilon_i$ replaced by 
the eigenvalues ${\cal E}_i$.
The results show a double-hump structure at the 
EP (Fig. 9 in \cite{mudiisro})
which corresponds exactly to the expression 
(\ref{sm5}) obtained analytically.  

Our conclusion from these results is, that nonlinear terms determine 
the resonance structure of the cross section in the 
neighborhood of an EP.

\section{S-matrix: resonance structure in the one-channel case}
\label{smatr1c}

According to textbooks, the resonance structure of the S-matrix is
well understood when all resonance states are coupled to one and the
same decay channel (this is the so-called {\it one-channel case}).
The resonance structure is calculated by means of the Hermitian 
formalism, in which no EPs are involved.
The basic results of theoretical and experimental studies agree under 
the condition that the resonances do not overlap, i.e. when they are 
well separated from one another in the cross section.

In order to receive a better understanding of the resonance structure 
of the cross section also in this simple case, we calculate it with
and without taking into account EM of the resonance states. In the 
first case ($\omega \ne 0$),  EPs are involved, while in the second 
case ($\omega = 0$) EPs do not appear. The Hamiltonian is
non-Hermitian in both cases. We compare the resonance structure of 
the S-matrix obtained in the two cases. 

We performed calculations with different values of $\omega$ and for 
different sets of resonance states. In Figs. \ref{fig3},
\ref{fig4} and \ref{fig5} 
we show typical results. They are obtained by choosing  
the values of $\omega$ 
to be the same as in Fig. \ref{fig1}  ($N=2$) and in 
Fig. \ref{fig2} left ($N=3$), respectively. The $\omega$ are complex 
and  near to those known from realistic systems, so that  
results can be obtained only numerically. 
The calculations are performed 
with the energies $\varepsilon_i \equiv e_i + \frac{i}{2} \gamma_i$
chosen in Figs. \ref{fig1} and \ref{fig2}, respectively. All states 
are coupled to one and the same continuum.

\subsection{Numerical results: resonance structure with $\omega \ne 0$}
\label{num6with}

\begin{figure}[ht]
\begin{center}
\includegraphics[width=9.cm,height=10.cm]{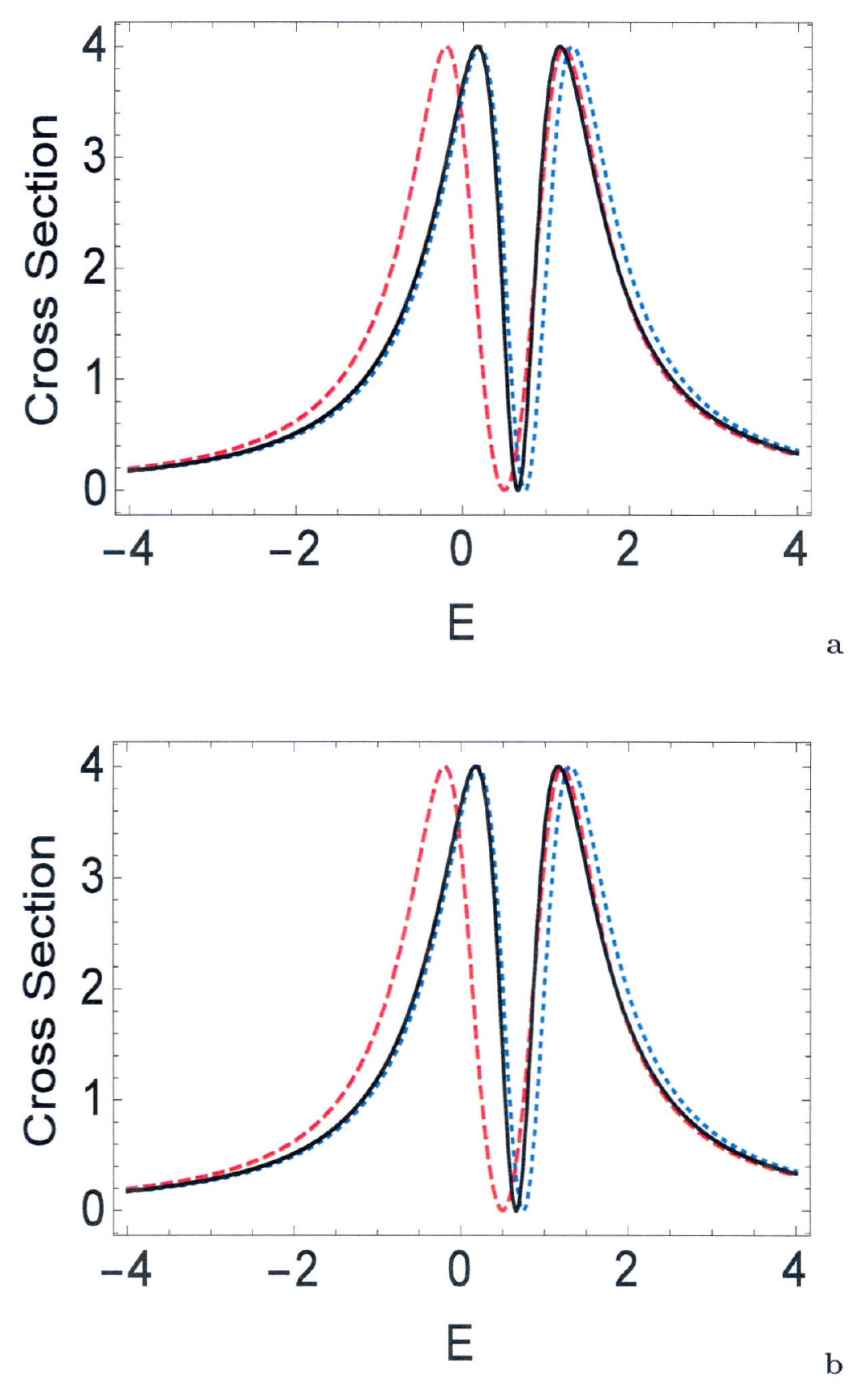}\\
\includegraphics[width=10.cm,height=7.cm]{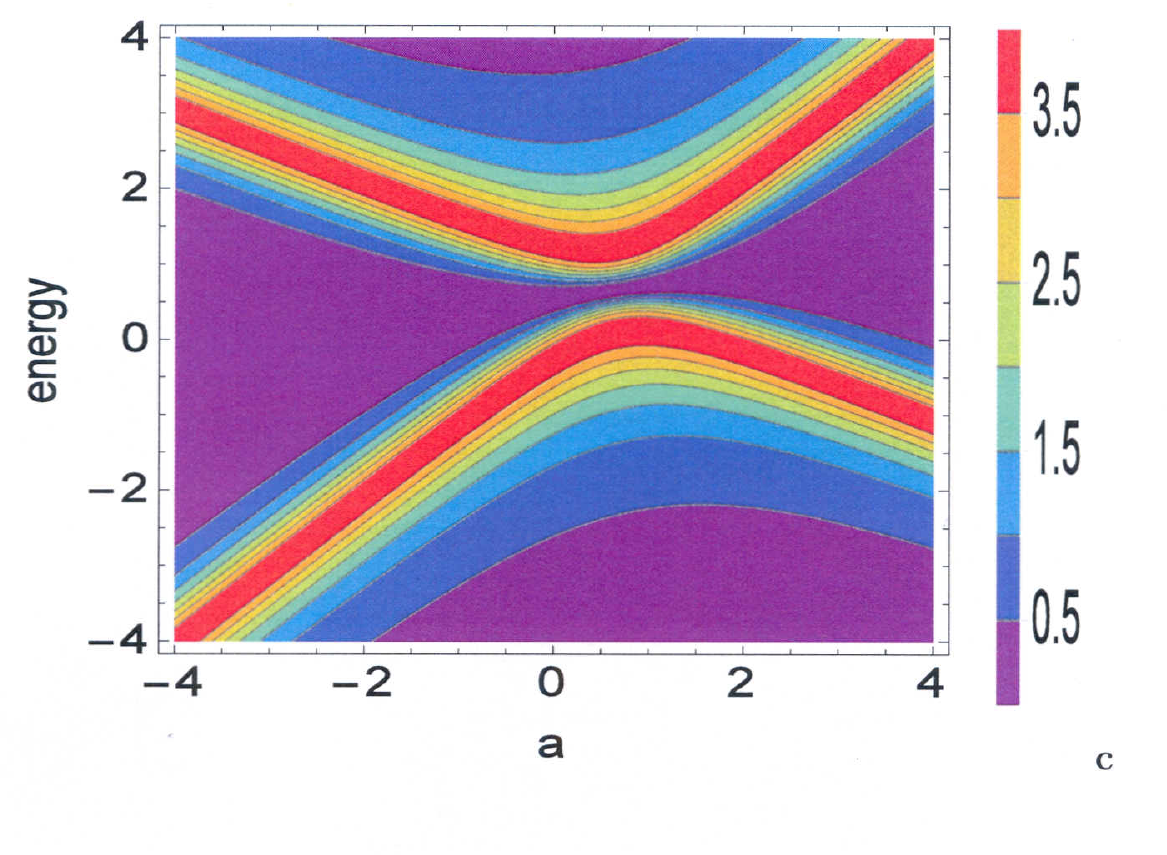}\\ 
\vspace*{-.7cm}
\end{center}
\caption{
\footnotesize{
Cross section with two resonance states. The parameters are the same
as in Fig. \ref{fig1} left, but $\omega = 0$ in (b). 
In (a) and (b),  $a=0$ (dashed red line); ~$a=1$
(dotted blue line); and $a=0.653333$ (full black line).
2D contour plot, $\omega = 0.01(i+\frac{1}{10})$ (c).  
}}
\label{fig3}
\end{figure}

\begin{figure}[ht]
\begin{center}
\includegraphics[width=9.cm,height=10.cm]{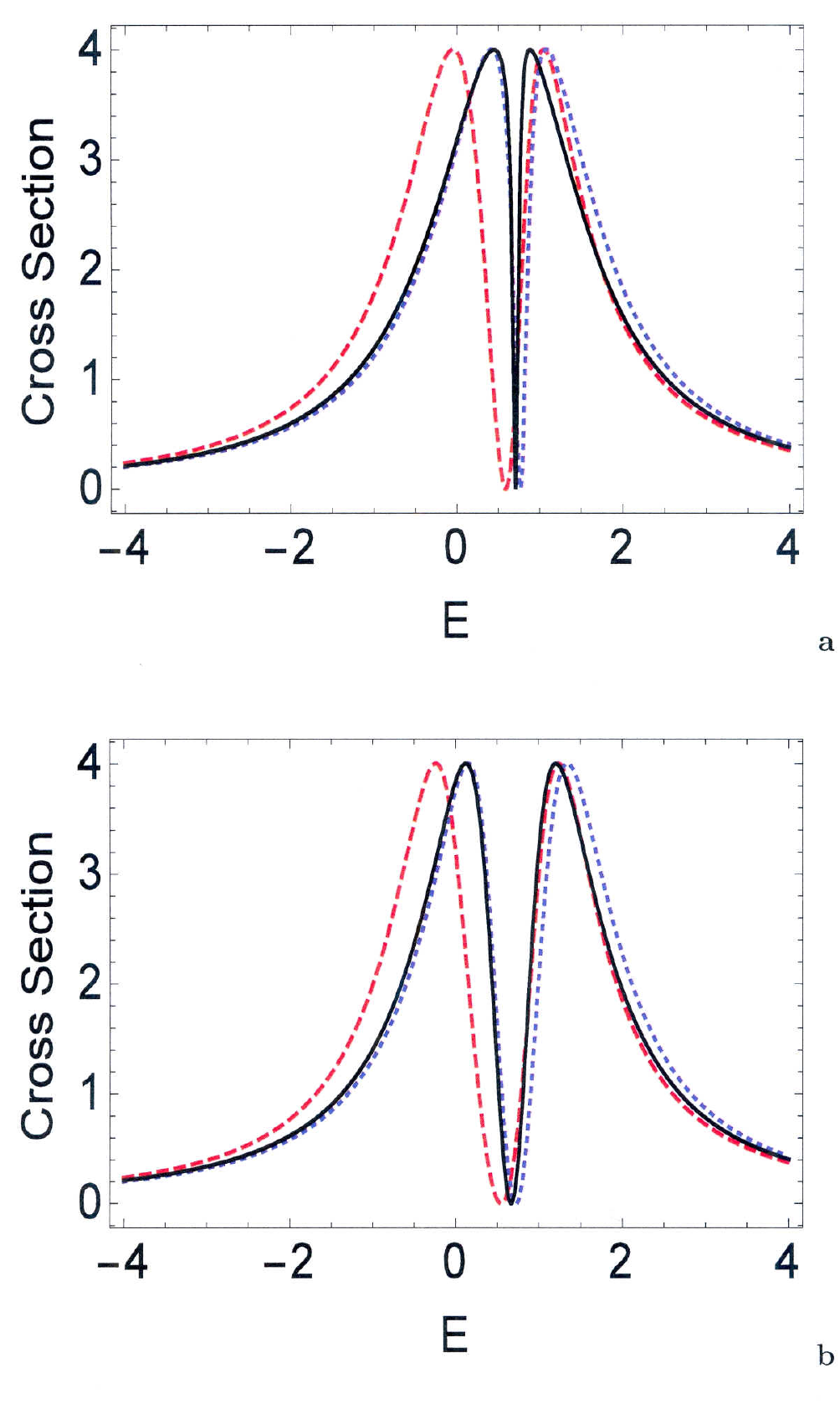} \\
\includegraphics[width=10.cm,height=7.cm]{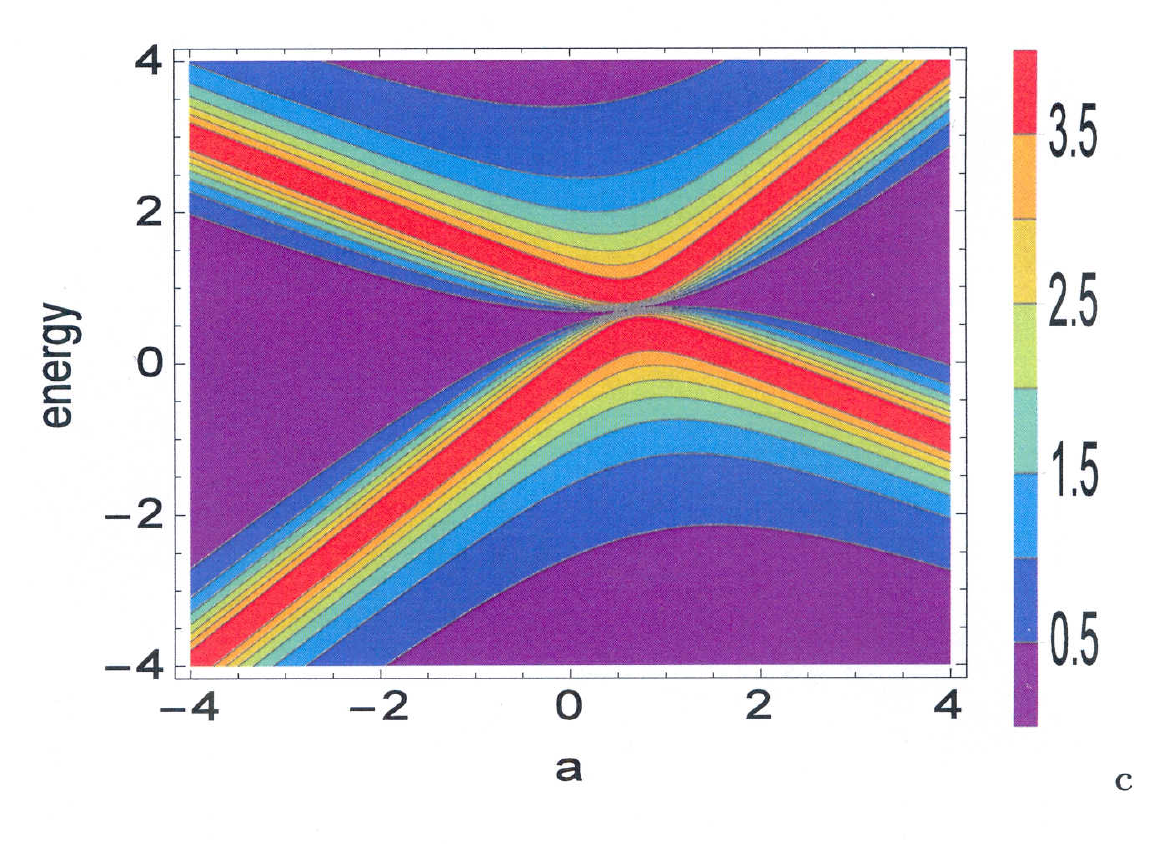}\\
\vspace*{-.7cm}
\end{center}
\caption{
\footnotesize{
Cross section with two resonance states. The parameters are the same
as in Fig. \ref{fig1} right, but $\omega = 0$ in (b). 
In (a) and (b),  $a=0$ (dashed red line); ~$a=1$
(dotted blue line); and $a=0.6502$ (full black line). 
2D contour plot,  $\omega = 0.5(i+\frac{1}{10})$ (c). 
}
}
\label{fig4}
\end{figure}

\begin{figure}[ht]
\begin{center}
\includegraphics[width=9.cm,height=10.cm]{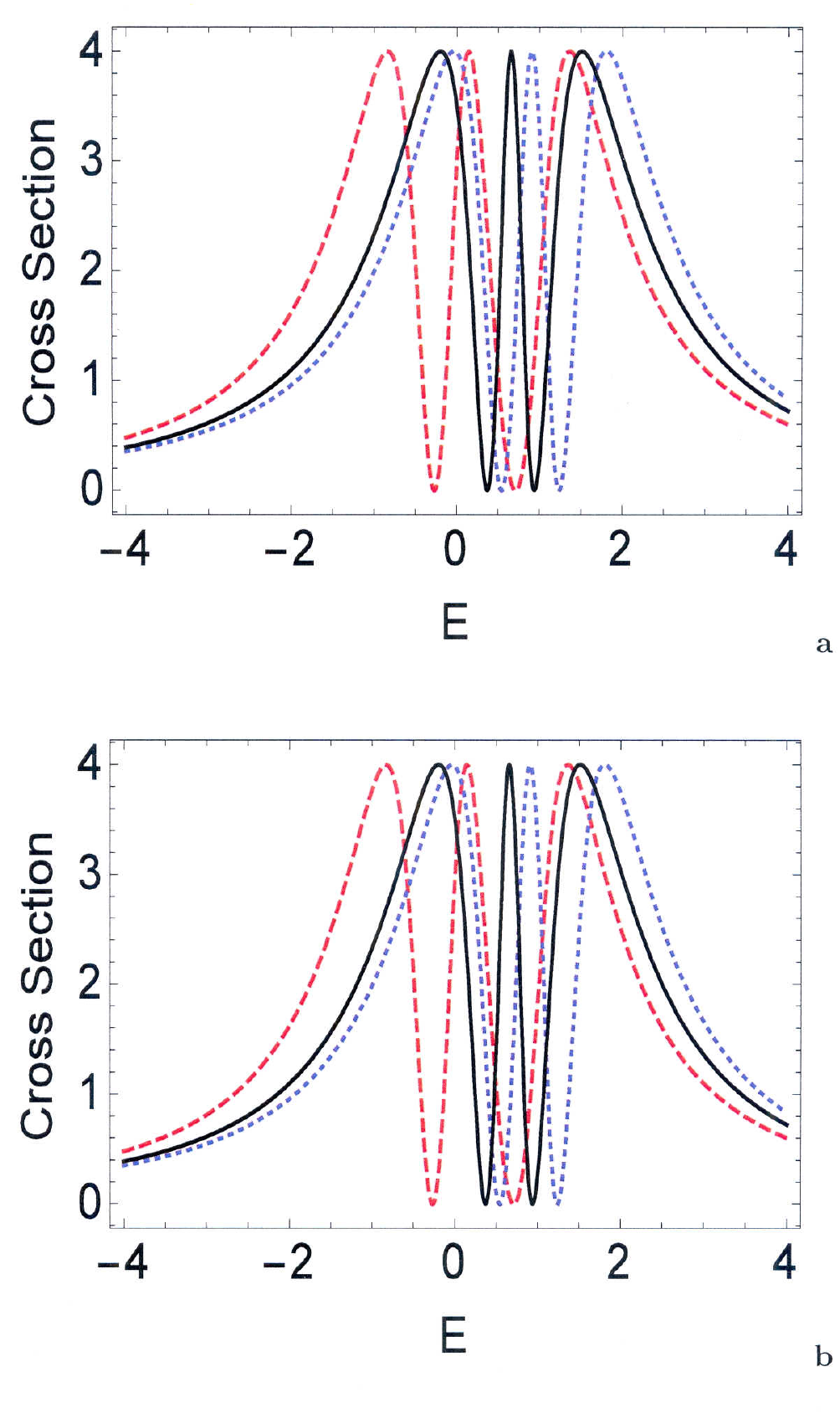} \\
\includegraphics[width=10.cm,height=7.cm]{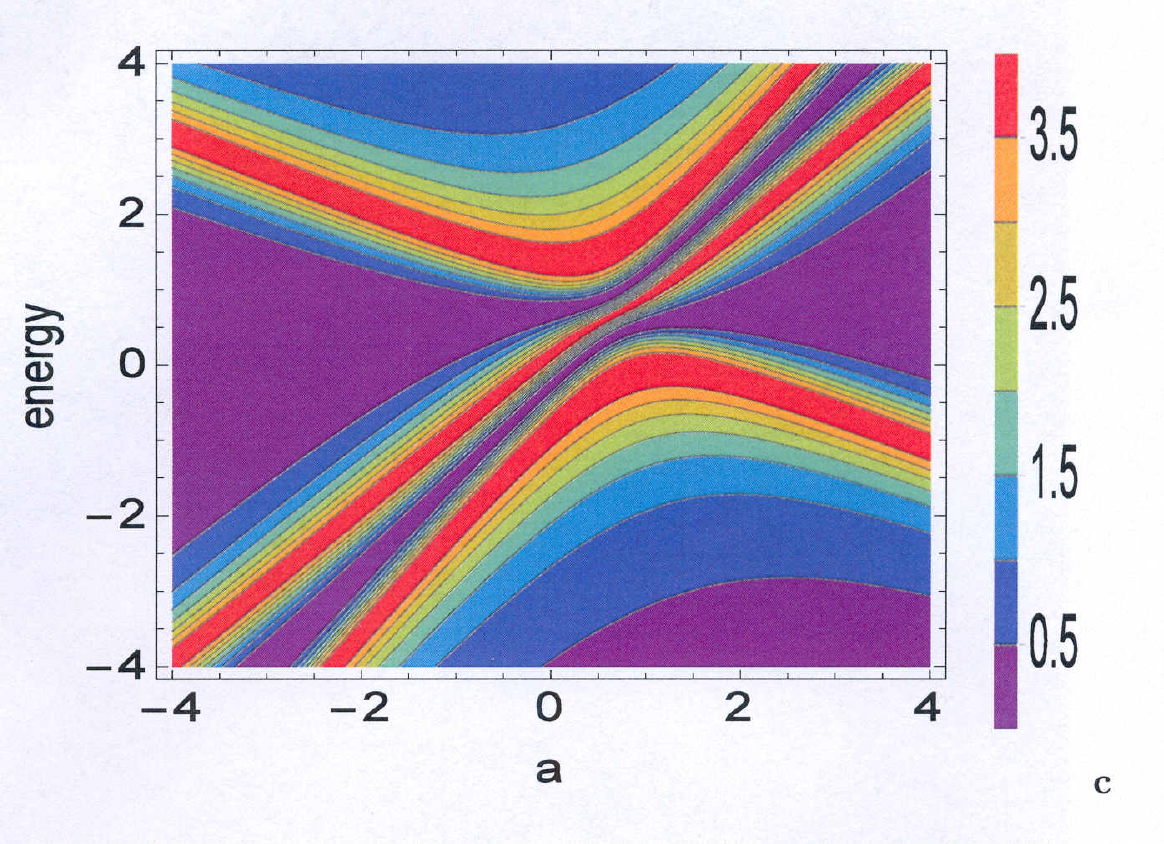}\\
\vspace*{-.7cm}
\end{center}
\caption{
\footnotesize{
Cross section with three resonance states. The parameters are the same
as in Fig. \ref{fig2} left, but $\omega = 0$ in (b). 
In (a) and (b),  $a=0$ (dashed red line); ~$a=1$
(dotted blue line); and $a=0.6502$ (full black line). 
2D contour plot, $\omega = 0$ (c). 
}
}
\label{fig5}
\end{figure}

Using  (\ref{cro}) and (\ref{sm4}) for the S-matrix, we calculated the
resonance structure of the cross section with, respectively, two and 
three resonance states under different conditions by  taking into account EM. 
In all cases with $N=2$ resonance states we see a double-hump structure, 
while the cross section shows a triple-hump structure when $N=3$. For 
examples see Figs. \ref{fig3}.a, \ref{fig4}.a and \ref{fig5}.a. 

In Fig. \ref{fig3}.a, the coupling of the states to the continuum is
relatively weak, see the  corresponding eigenvalue 
pictures Fig. \ref{fig1}.a,b. The resonance part of the S-matrix shows 
the typical two-hump structure. 

In Fig. \ref{fig4}.a  the difference between the two widths
$\gamma_i/2$ is relatively large, see the corresponding eigenvalue 
pictures Fig. \ref{fig1}.f,g. 
In order to see the influence of an EP, also the coupling
strength $|\omega| $ has to be relatively large in this case.   
According to Fig. \ref{fig1}.g, it is
$\sum_{i=1,2} \Gamma_i \approx \Gamma_2$ where $\Gamma_2$ is the width 
of the short-lived state $2$. As can be seen from Fig. \ref{fig4}.a, the
cross section shows the characteristic  double-hump structure not only
in the very neighborhood of the EP but also beyond this value.

Using Eqs. (\ref{cro}) and (\ref{sm4}) for the S-matrix,
we are able to reproduce the double-hump structure of the cross section 
as a function of the coupling strength $\omega$ which is shown in
Fig. 9 of the old paper \cite{mudiisro}.  The calculations 
in \cite{mudiisro} are performed  on
the basis of the standard S-matrix theory,
however with the energies $\varepsilon_i$ replaced by the eigenvalues 
${\cal E}_i$. The role of the interference of the  
different contributions to the resonance structure is also
shown in Fig. 9 in \cite{mudiisro}.
 
In  Fig. \ref{fig5}.a, we show the results with $N=3$ for the case that
the widths of the three states are similar to one another
and $\omega$ is relatively small, see the
corresponding eigenvalue pictures Figs. \ref{fig2}.a,b. 
The cross section shows the
typical three-hump structure at different values of the parameter 
$a$ near to the region  with several neighbored EPs
as well as beyond it.  

The 2D-contour plots of the resonance structure of the cross section
with two levels, calculated  with $\omega \ne 0$, are shown in 
Figs. \ref{fig3}.c and  \ref{fig4}.c.
In both cases, the cross section falls down steeply to its minimum  
value between the two EPs. 
Here the eigenfunctions are (almost) orthogonal and mixed 
in the set of basic wavefunctions $\{\Phi_n^0\}$, see
Fig. \ref{fig1}. As can be seen from Fig. \ref{fig4}.c, the minimum
value appears at the value of maximum width bifurcation (see
Fig. \ref{fig1} right column).

The 2D-contour plot of the resonance structure of the cross section
with three levels, calculated  with $\omega \ne 0$, is the same as
that calculated with  $\omega = 0$ (see next section 
\ref{num6without}).

In any case, the 2D-contour plots of the resonance structure 
of the cross section should not be confused with eigenvalue 
trajectories that avoid crossing. Furthermore, they are not 
symmetric with respect to $E=0$ corresponding to the
eigenvalue pictures Figs. \ref{fig1} and \ref{fig2} (in difference to
Fig. 9 for the cross section in \cite{mudiisro} that is related to 
eigenvalue figures which are symmetric with respect to $E=0$).

\subsection{Numerical results: resonance structure with $\omega = 0$}
\label{num6without}

We compare the resonance structure of the S-matrix 
obtained in the non-Hermitian formalism with taking into account EM 
($\omega \ne 0$) to that obtained without EM
(corresponding to $\omega =0$).
Typical results are shown  in Figs. \ref{fig3}.b and \ref{fig4}.b
for $N=2$ and in Fig. \ref{fig5}.b. for $N=3$. These figures have to be 
compared with, respectively, Figs. \ref{fig3}.a, \ref{fig4}.a
and \ref{fig5}.a.

As in all our calculations, the resonance
structure of the S-matrix is almost the same for $\omega = 0$ and
$\omega \ne 0$. Differences in the resonance structure of the S-matrix
can be seen only when $\omega$ is large, see Fig. \ref{fig4}.a as
compared to Fig. \ref{fig4}.b. In all other cases, the resonance
structure is typically the same, see for example Fig. \ref{fig3}.a 
as compared to Fig. \ref{fig3}.b and Fig. \ref{fig5}.a 
as compared to Fig. \ref{fig5}.b.

In Fig.  \ref{fig5}.c we 
show the 2D-contour plot of the cross section with three resonances,
calculated with $\omega = 0$. It looks like that obtained with 
$\omega \ne 0$. Also the results for $N=2$  with $\omega = 0$ 
are typically the same as those with $\omega \ne 0$ (which are shown 
in Figs. \ref{fig3}.c and \ref{fig4}.c).

In the 2D-contour plot of the resonance structure with three levels  
(Fig.  \ref{fig5}.c), we see two second-order EPs instead of a
third-order EP. This result corresponds to the discussion on
third-order EPs in Sect. \ref{third}.

\subsection{Influence of exceptional points}
\label{infl}

In Figs. \ref{fig3}, \ref{fig4} and \ref{fig5} we have shown the
numerical results obtained for the resonance structure of the 
S-matrix when the system is considered, respectively, with EM 
($\omega \ne 0$) and without EM ($\omega = 0$).
We considered the most sensitive situation where the resonance
structure is influenced by two adjoining EPs. 
EPs and EM appear only when $\omega \ne 0$. 
Nevertheless, the resonance structure of the 
cross section is almost  the same in the two cases.
 
This result is valid not only when the number of
resonances is two but also when it is larger  than two. 
This means that the resonance structure of the cross section is almost 
 independent of EM, i.e. on the  
coupling of the states via one common continuum of scattering states.
  
This unexpected result can be explained in the following manner. 
The evolution of the system between the two EPs is driven exclusively by the 
nonlinear source term 
of the Schr\"odinger equation (\ref{eif11}) since $\omega$ is constant 
in our calculations and can therefore not be responsible for the 
width bifurcation. Obviously, the
nonlinear source term is able, in the one-channel case, to largely 
conserve the resonance structure of the cross section. 

Altogether, we have here some type of self-affirmation. Analytical 
results for the resonance structure of the cross section 
can be obtained, in the one-channel case (with well separated
resonances), when the system is described by a Hermitian
operator  the eigenvalues and  eigenfunctions 
of which are smoothly parameter dependent. These results agree 
quite well with those of experimental observations. 
The description of  the system as a closed system seems therefore 
to be justified.  In addition,
more complicated cases with e.g. more than one open channel, cannot 
be solved analytically in the standard theory. Thus, 
the justification of the Hermitian approach for the description 
of the system (with well separated resonances) rests solely 
on the analytical results obtained for the one-channel case.   

Our results for the one-channel case show that this case cannot 
be used 
in order to prove or disprove the Hermitian quantum physics.
To that purpose, the study of 
more complicated cases is needed, see the 
next section \ref{smatr2c}.

We mention here that  the resonance scattering at third-order EPs is
studied in \cite{heiwu} by using a method that is different from
ours. Also in these calculations, 
three peaks appear in the cross section. According to \cite{heiwu}, 
the ``sprouting out'' of the
three levels under parameter variation depends on the particular
parameter chosen. A similar result is obtained   
\cite{epj2,pra93}  in the framework of the formalism presented 
in the present paper. However, a third-order EP does not appear in our 
calculations, see Fig. \ref{fig5}.c. Instead we see several 
second-order EPs and hints to them, respectively, what agrees 
with the discussion in Sect. \ref{third}.

\section{S-matrix:  resonance structure in the two-channel case}
\label{smatr2c}

We will not provide here new numerical results
for the two-channel case. Instead we refer to results obtained 
a few years ago \cite{muro,burosa1,burosa2} for the 
transmission through a small system (quantum dot).
In order to describe transmission, we have to consider at least 
two channels: the entrance and the exit channel.  In 
\cite{muro}, unexpected experimental results 
\cite{yacobi,schuster,heiblum} on the resonance structure of the
transmission could be explained. In the papers
\cite{burosa1,burosa2}, different
calculations are performed for both, a system with a small number of 
resonance states as well as for a system with many states. In the 
last case, the calculations are performed 
first in the tight-binding approach according to the formalism
presented by Datta \cite{datta}.
Then, the non-Hermitian Hamilton operator is diagonalized and the 
eigenvalues and eigenfunctions are determined. This formalism is 
equivalent to that used in the present paper (see
p. 437 in \cite{knoss}). 
The calculations for systems with a small number of states are
performed according S-matrix theory and using 
tight-binding approach \cite{saburo}. 

The results for the two-channel case
are more interesting than those for the one-channel 
case discussed in Sect. \ref{smatr1c} since the influence of EPs and 
EM can be seen immediately \cite{muro,burosa1,burosa2}. 
The resonance structure of the transmission 
can be traced back to the eigenvalues of the non-Hermitian operator;  
EM of the states can generally not be neglected;
and the phase rigidity is anticorrelated with the transmission 
probability. The last property is the most interesting one. It 
has no analog in the standard formalism. 

The  phase rigidity is a theoretical value
characteristic of the non-Hermitian formalism. It  can be traced 
experimentally in a microwave 
billiard \cite{richter2}. It will however be difficult to study 
it directly  in a realistic system. According to 
the above mentioned  numerical results \cite{burosa1,burosa2} it
is, however, anticorrelated with an observable value, namely with 
the transmission probability. 

This anticorrelation of the theoretical value (phase rigidity) 
with an observable one (transmission probability)   
allows us  to really test the  non-Hermitian formalism. 
Moreover, when this anticorrelation 
really exists, it is of high interest for applications.

\section{Discussion of the results and summary}
\label{disc}

A critical consideration of the standard Hermitian formalism 
for the description of open quantum systems is possible 
by starting from a general non-Hermitian formalism
\cite{comment3} which includes 
the Hermitian quantum physics as a limiting case. In the 
non-Hermitian formalism, the normalization of the eigenfunctions of 
the Hamiltonian can freely be chosen \cite{gurosa}. 
If it is chosen by means of (\ref{eif3}), 
the non-Hermitian quantum formalism fulfills the condition to  
approach, on the one hand, the standard Hermitian quantum physics
under certain conditions (that can be formulated) and to be, 
on the other hand, more general than it.  The mathematical 
consequences are the following.
\vspace{-1cm}
\begin{verse}
\item
The phases of the eigenfunctions
relative to one another are not rigid, see Sect. \ref{eigfu}.
This fact agrees with the basic relation (\ref{sec8}) which is valid 
in approaching an EP: the two eigenfunctions 
$\Phi_1$ and $\Phi_2$ of ${\cal H}^{(2)}$
 are (almost) orthogonal to one another when the two eigenstates 
$1$ and $2$  are distant from one another, while the orthogonality is 
completely lost in approaching an EP.

\item
The eigenstates contain EM and differ therefore from 
the original eigenstates. A mixing of the wavefunctions of 
only two states may appear at low level density
such that it may be difficult to choose the basic set of 
"pure" wavefunctions. At high level density EM causes a dynamical
phase transition
which is non-adiabatic due to the involved nonlinear processes
\cite{epj2}.

\item
Some well-known unsolved puzzles of standard Hermitian 
quantum physics do not appear in the non-Hermitian description 
of open quantum systems \cite{comment3}. Among others, the problem of the 
Schr\"odinger cat and the short tunneling
time characterizing  the decay of the states in the 
Hermitian quantum 
physics, are not puzzling when the system is considered to be open. 
Furthermore, the nonlinear processes involved in the 
non-Hermitian formalism, are irreversible, see the discussion around
Fig. 9 in  \cite{ropp}. 

\end{verse}

In Sects. \ref{eigf2}, \ref{eigf3} and \ref{smatr1c} of the present 
paper, we have shown numerical results obtained in the framework of 
non-Hermitian quantum theory for a system that is 
coupled to one common channel \cite{comment3}. 
We consider systems with $N=2$ and 
$N=3$  states in the most sensitive parameter range in which
the dynamics of the system is determined by two EPs.
We have compared the results, obtained for the same situation,
with and without taking into account EM. In the 
first case, EPs are caused by the EM of the states. In the 
second case, however, neither EPs nor EM appear. Nevertheless, 
the resonance structure of the S-matrix is almost the same in the two 
cases. This result does not depend on the number of states 
taken into account in the calculation. 

In our calculations with non-Hermitian Hamiltonian, the coupling 
strength $\omega$ between system and environment is chosen to be
fixed. Width 
bifurcation of the states may be caused therefore exclusively by the 
nonlinear terms contained in 
the Schr\"odinger equation at and near to an EP. 
These nonlinear terms conserve, obviously, the resonance structure 
of the cross section in the one-channel case. 
Thus, the one-channel case does not allow us to test  the 
non-Hermitian formalism.

The situation is completely different when the system is coupled 
to two (or more) channels. A prominent example is the transmission 
through, e.g., a quantum dot. Here, at least two different channels 
are involved: entrance and exit channel. In the present 
paper, we do not provide new numerical results. Instead we refer to 
some results obtained earlier  (see Sect. \ref{smatr2c}).  
Most interesting is the anticorrelation between phase rigidity and 
transmission probability which can be seen clearly in the results of 
different calculations.

Thus, the observation of non-analytical effects in the transmission 
through a quantum dot is not in contradiction to the 
results known from the standard S-matrix description in the
one-channel case. Quite the contrary,
these effects are characteristic of the non-Hermitian theory of 
open quantum systems. They exist also in the one-channel case 
where they can, however, not be seen due to their
suppression by the nonlinear terms of the Schr\"odinger equation
near to EPs.

\section{Conclusions}
\label{concl}

The results of the present paper  answer the
questions asked in the Introduction.
Although EPs influence the dynamics of open quantum systems, they
cannot be observed directly.  In the one-channel case, 
the resonance structure of the cross section can be described well 
{\it without} taking them into account. The reason for this 
unexpected result  
are the nonlinear processes caused by the EPs. They restore, in
the one-channel case, the original resonance structure of the cross 
section and hide the influence of the EPs onto observable values.  
A cursory consideration allows therefore 
the conclusion that EPs do not play any role in open quantum systems. 

This conclusion can be justified however only in the one-channel
case. The two-channel case is much richer and more interesting. 
In the Introduction, we pointed to the so-called phase lapses
observed experimentally in the transmission through a quantum dot. 
These unexpected results are explained by means of the existence
of EPs. We mentioned moreover in Sect. \ref{smatr2c}  the  
results obtained theoretically for the transmission through a
localized quantum system by using the tight-binding approach. 
We underline however that, in any case, the results of non-Hermitian
quantum physics differ from those of Hermitian quantum physics
only little in a parameter range that is {\it not} influenced
by EPs. In this parameter range, Fermi's golden rule holds.

The results for the two-channel case show, under different conditions, 
an anti-correlation between phase
rigidity and transmission probability, i.e. between an {\it internal}
property of the eigenfunctions of the non-Hermitian Hamilton operator
and  an observable value. Based on the non-Hermitian quantum
theory formulated in the present paper, 
an experimental test of this relation will
contribute, among others, also to an understanding of the short
tunneling time \cite{tunnel} that is observed experimentally.    

The transmission through a small system needs to be studied 
in future in more detail, theoretically as well as experimentally. 
On the one hand, it allows us to test the non-Hermitian quantum 
theory for open quantum systems \cite{comment3},
since it relates the theoretical value phase rigidity to 
the observable value of the transmission probability.  
On the other hand, 
the anticorrelation between these two values will open the door   
to important applications.


\begin{thebibliography}{99}
\bibitem{comment2} The coalescence of two eigenvalues of a
  non-Hermitian operator should not be confused with the degeneration
  of two eigenstates of a Hermitian operator. The eigenfunctions of
  two degenerate states are different and orthogonal while those of
  two coalescing states are biorthogonal and differ only by a phase,
  see Eq. (\ref{sec8}).
\end{thebibliography}
\end{document}